\documentstyle[twocolumn,aps]{revtex}
\input epsf
\begin{document}
\draft
\title{Statistical Properties of Statistical Ensembles of Stock Returns}
\author{Fabrizio Lillo and Rosario N. Mantegna}
\address{
Istituto Nazionale per la Fisica della Materia, Unit\`a di Palermo\\
and\\
Dipartimento di Energetica ed Applicazioni di Fisica,
Universit\`a di Palermo, Viale delle Scienze, I-90128,
Palermo, Italia}
\date{\today}
\maketitle
%\receipt{}
%----------------------------------------------------------------------
\begin{abstract}
We select $n$ stocks traded in the New York Stock Exchange and we 
form a statistical ensemble of daily stock returns for each of the 
$k$ trading days of our database from the stock price time series. 
We analyze each ensemble of stock returns by extracting its first 
four central moments. We observe that these moments are fluctuating 
in time and are stochastic processes themselves. We characterize 
the statistical properties of central moments by investigating 
their probability density function and temporal correlation properties. 
\end{abstract}
%----------------------------------------------------------------------
%\pacs{0X.XX.+x,YY.YY.zz}
%\newpage

\narrowtext

\bigskip

In the last years a large amount of statistical analyses of the dynamics of the price time series of a {\it single} stock  
has been performed by physicists interested in the modeling of financial markets \cite{Palermo}. In this paper we present the results of an empirical analysis performed by taking a different approach. We investigate the statistical properties of daily returns of $n$ selected stocks simultaneously traded in a financial market.
There are two main motivations for this kind of analysis. From a fundamental point of view our analysis may help in understanding collective behaviors in stock markets. These behaviors become of great importance in times of financial turmoil when stocks in the market become more interlinked, and during market crashes. From an applied point of view our analysis may be useful in the management of large portfolio of stocks.

The investigated market is the New York Stock Exchange (NYSE) during the 12-year period January 1987 to April 1999 comprising 3113 trading days. We select four ensembles of $n$ stocks. The number of stocks in each ensemble is not always constant during the investigated period because the number of stocks is rapidly increasing in the NYSE, ranging from approximately $1100$ in 1987 to approximately $2800$ in 1999. Old stocks disappear and new ones start to be traded in the market. Moreover for each ensemble we consider only the stocks traded in the NYSE and we exclude those traded in the NASDAQ or AMEX market. Hence $n$ is constant or slowly increasing with time in the selected ensembles of stocks: (i) $n=30$ stocks which are used to compute the Dow Jones Industrial Average (DJIA30); (ii) $n>86$ stocks which are used to compute the Standard \& Poors 100 Index (SP100); (iii) $n>313$ stocks which are used to compute the Standard \& Poors 500 Index (SP500), and (iv) all the $n>1100$ stocks traded in NYSE (NYSE). 
The variable investigated in our analysis is the daily return, which is defined as
\begin{equation}
R_i(t+1)\equiv\frac{Y_i(t+1)-Y_i(t)}{Y_i(t)},
\end{equation}  
where $Y_i(t)$ is the closure price of $i-$th stock at day $t$. For each day we consider $n$ returns. $n$ is about $30, 90, 420, 2100$ depending on the chosen set. 
A first analysis concerns the distribution of returns at a given day $t$. We observe that in many days the central part of this distribution is approximated by a Laplace or double exponential distribution \cite{Papoulis}. Significant changes in the shape and scale are frequently observed, especially in times of financial turmoil \cite{Lillo}. Laplace distribution has been considered recently in economic analysis of the growth dynamics of companies \cite{Stanley96}.
In order to characterize more quantitatively the return distribution at day $t$, we determine the first four central moments for each of the $3113$ trading days of the 4 sets of stocks considered. Specifically, we consider the mean, the standard deviation, the skewness and the kurtosis defined as
\begin{eqnarray}
\mu (t)=\frac{1}{n}\sum_{i=1}^n R_i(t), \\ \sigma (t)= \sqrt{\frac{1}{n-1}\sum_{i=1}^n(R_i(t)-\mu(t))^2}, 
\\
\rho (t) = \frac{1}{n} \sum_{i=1}^n \left(\frac{R_i(t)-\mu(t)}{\sigma(t)}\right)^3, \\ \kappa (t) = \frac{1}{n} \sum_{i=1}^n\left(\frac{R_i(t)-\mu(t)}{\sigma(t)}\right)^4.
\end{eqnarray}
The mean $\mu(t)$ gives a measure of the general trend of the market at day $t$. The standard deviation $\sigma(t)$ controls the width of the distribution and gives a measure of the {\it variety} of behaviors observed in the financial market. A large value of $\sigma(t)$ indicates that different companies show very different behaviors at day $t$. Skewness $\rho(t)$ and kurtosis $\kappa(t)$ are scale-free parameters, whose values depend on the shape of the distribution but not on its  scale. $\rho(t)$ describes the asymmetry of the distribution with respect to $\mu(t)$. A positive value of $\rho(t)$ indicates that few companies perform great profits, and many companies have small losses at day $t$ with respect to the mean. A negative value of $\rho(t)$ corresponds to the complementary case. Finally the kurtosis $\kappa(t)$ gives a measure of the distance of the distribution from a Gaussian distribution. 
In our analysis we have discarded returns whose absolute value was $|R_i(t)|>0.5$, because some of these returns might

\begin{figure}[t]
\epsfxsize=3in
\epsfbox{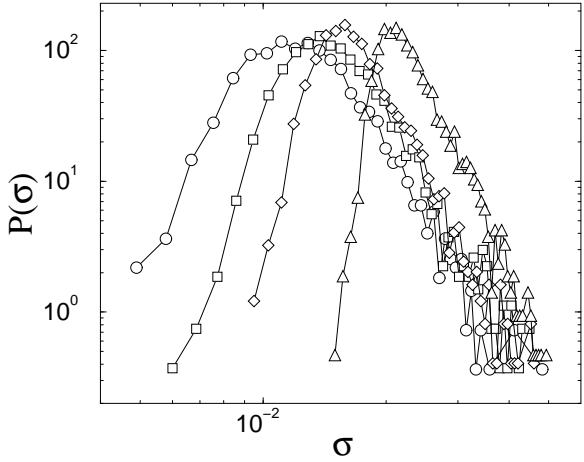}
\caption{Probability density function of the standard deviation (variety) $\sigma(t)$ for the four ensembles considered: DJIA30 (circle), SP100 (square), SP500 (diamond) and NYSE (triangle).}
\label{fig1}
 \end{figure}

 be attributed to errors in the database. Similar errors would strongly affect the estimation of higher moments because statistical analyses of moments of a distribution 
higher than the second are more and more sensible to extreme values.

We obtain the values of the four moments for each trading day. These quantities are not constant but fluctuate in time. By observing the time evolution of $\mu(t)$, we note that several trading days are present in which big jumps of the mean are observed. These findings can be evaluated more quantitatively by investigating the empirical probability density function (PDF) of $\mu(t)$ temporal series. This PDF is also approximated by a Laplace distribution. 

In Fig. 1 we show the PDFs of the variety $\sigma(t)$. We observe that the distribution is roughly log-normal with a approximately power-law tail observed for the higher values for each ensemble considered. We note that distributions do not coincide for different ensembles. In particular the mean of $\sigma(t)$ increases by increasing the number of stocks considered and by decreasing the (average) capitalization of the stocks. Indeed the stocks which compose the DJIA30 set have great capitalization and small volatility. On the other hand in the NYSE set there are present companies with both small and large capitalization and with different levels of volatility. The NYSE set is more heterogeneous than DJIA30 set and this is reflected in a greater value of variety $\sigma(t)$.

The higher moments are extremely sensible to rare events. The PDF of the skewness is non-Gaussian with fat tails and is slightly asymmetrical around the value $\rho=0$ (especially for NYSE set). Positive values of the skewness are a bit more probable than negative ones. This implies that days in which few companies reach great gains and many companies have small losses with respect to the mean are slightly more frequent than the complementary case.
The PDFs of the kurtosis $P(\kappa)$ are approximately characterized by a power-law tail $\kappa^{-\gamma}$ for higher values of $\kappa$ for the four ensembles of stocks. The exponent $\gamma$ of the power-law region is approximately 
equal to $2$ and becomes slightly greater moving from the NYSE to the DJIA30 sets.   
In summary the first four central moments of the distribution of daily returns are distributed in a non-trivial way.

\begin{figure}[t]
\epsfxsize=3in
\epsfbox{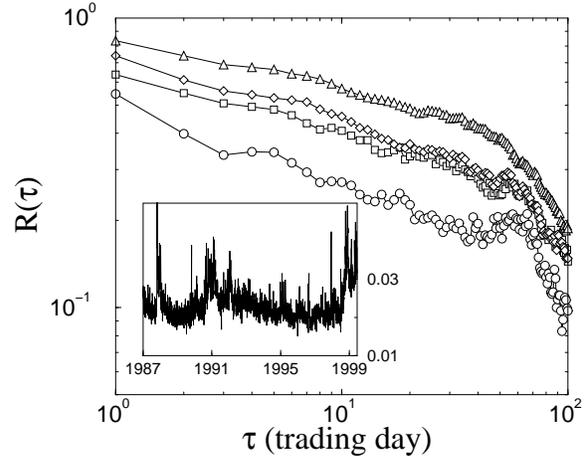}
\caption{Log-log plot of the autocorrelation function $R(\tau)$ of the variety $\sigma (t)$ as a function of the time lag $\tau$ (in trading days) for the four ensembles considered: DJIA30 (circle), SP100 (square), SP500 (diamond) and NYSE (triangle). In the inset we show the time evolution of the variety for NYSE ensemble. On 19 October 1987 variety reached the value $\sigma=0.096$ out of scale in the inset.}
\label{fig2}
 \end{figure}

In order to better characterize the temporal evolutions of $\mu(t)$ and $\sigma(t)$, we investigate their memory properties. To this end we calculate their autocorrelation functions. We find that the mean is delta correlated, whereas a different behavior is observed for $\sigma(t)$. Fig. 2 shows the autocorrelation function of the variety $\sigma(t)$ for the four ensembles considered in a log-log plot. We observe that the autocorrelation function of empirical data is well approximated by a power-law function $R(\tau)\propto \tau^{-\delta}$. By performing a best fit of $R(\tau)$ with a maximum time lag of $50$ trading days, we determine $\delta$ as $0.27$ (DJIA30), $0.25$ (SP100), $0.26$ (SP500) and $0.20$ (NYSE). These results indicate that a long-time memory is present in the market for the variety $\sigma (t)$. We observe a power-law autocorrelation function also for the quantity $|\mu(t)|$.
We mention that the behaviors of the autocorrelation function of $\mu(t)$ and $\sigma(t)$ are consistent with the results of our analysis of their Fourier transform. We observe that $\mu(t)$ has a white noise power spectrum, whereas the variety $\sigma(t)$ has a power-law power spectrum.  

In conclusion we have introduced the concept of variety of a statistical ensemble of stocks traded in a financial market. Statistical properties of variety are non-trivial and are characterized by a non-Gaussian PDF and by a long-term time memory.  

The authors thank INFM and MURST for financial support.
%----------------------------------------------------------------------
%\newpage
%\centerline{\bf References}

%----------------------------------------------------------------------

%
%

\end{document}